# Coupling between electronic and structural degrees of freedom in the triangular lattice conductor $Na_xCoO_2$

Q. Huang,[1] M.L. Foo,[2] R.A. Pascal Jr.,[2] J.W. Lynn,[1] B.H. Toby,[1] Tao He,[3] H.W. Zandbergen,[4] and R.J. Cava[2]

[1] NIST Center for Neutron Research, NIST, Gaithersburg, MD 20899
[2] Department of Chemistry, Princeton University, Princeton NJ 08540
[3] DuPont Central Research and Development Experimental Station, Wilmington, DE 19880
[4] National Centre for HREM, Department of Nanoscience, Delft Institute of Technology, Al Delft, The Netherlands

The determination by powder neutron diffraction of the ambient temperature crystal structures of compounds in the $Na_xCoO_2$ family, for $0.3 < x \leq 1.0$, is reported. The structures consist of triangular $CoO_2$ layers with Na ions distributed in intervening charge reservoir layers. The shapes of the $CoO_6$ octahedra that make up the $CoO_2$ layers are found to be critically dependent on the electron count and on the distribution of the Na ions in the intervening layers, where two types of Na sites are available. Correlation of the shapes of cobalt-oxygen octahedra, the Na ion positions, and the electronic phase diagram in $Na_xCoO_2$ is made, showing how structural and electronic degrees of freedom can be coupled in electrically conducting triangular lattice systems.

**Introduction**

Square planes of transition metals and oxygen are the basis for scientifically and technologically important materials such as perovskite ferroelectrics, superconductors and ferromagnets. Compounds based on triangular arrangements of transition metals and oxygen are much less generally studied, except in the context of magnetism in electrically insulating materials, where the placement of antiferromagnetically coupled magnetic atoms in triangle-based lattices leads to frustration of the low temperature magnetic state of the system due to conflicting near neighbor spin-spin interactions (1). The introduction of charge carriers into such triangular magnetic lattices in sufficient number to yield metallic conductivity has been a goal of research for some time. $Na_xCoO_2$, with a crystal structure consisting of alternating layers of Na ions and triangular $CoO_2$ planes, is the embodiment of such a system. The charge on the $CoO_2$ planes is controlled by the degree of filling of the Na layer – in a manner fully analogous to the relationship between the charge reservoir layers and the $CuO_2$ planes in the high $T_c$ superconductors (2). The properties of $Na_xCoO_2$ are not disappointing: a high thermoelectric power metallic conductor is found for $Na_{0.7}CoO_2$ (3, 4), superconductivity is induced in $Na_{0.35}CoO_2$ only when it is intercalated with water (5), and $Na_{0.5}CoO_2$ appears to be insulating at low temperatures due to the presence of ordering of Co ion charges (6). Oxide materials based on square-based transition metal-oxygen lattices display a broad range of effects due to the coupling of electronic, magnetic, and structural degrees of freedom, but until now there has been no model system for investigating such effects in triangle-based electronic conductors. Here we report structural studies of the average crystal structure of $Na_xCoO_2$ over a wide range of Na ion filling of the charge reservoir layer. We find that changes in the electron count strongly affect the structure of the $CoO_2$ plane. The results suggest that conducting triangular lattice systems display their own class of structural and electronic coupling phenomena distinct from those previously studied in square-based systems.

Hexagonal $Na_xCoO_2$, which has two triangular $CoO_2$ layers per unit cell, has been reported to exist for $0.25<x<\sim0.8$ (6-13). It exhibits metallic behavior for a wide range of x, with the exception of $x = 1/2$, where an insulating state is found at low temperatures (6). Magnetic ordering has been reported for compositions near $Na_{0.75}CoO_2$ (7-9). Na contents lower than 0.7 in $Na_xCoO_2$ can only be obtained by room temperature deintercalation of sodium from $Na_{0.7}CoO_2$, and in one case crystals grown by the floating zone technique have been reported to have $x > 0.75$ (9). Crystal structures have been reported for some $Na_xCoO_2$ compositions (10-13), and $Na_{0.3}CoO_2$ has recently been studied by neutron diffraction due to its importance as the host for the superconducting phase (14, 15). An electron diffraction study of the $Na_xCoO_2$ series (16) has revealed a complex series of ordered Na distributions within the Na planes at different Na compositions.

This report describes the results of crystal structure determinations by powder neutron diffraction of a series of samples with x between 0.3 and 1.0 in $Na_xCoO_2$, with an eye toward interpreting the electronic phase diagram. Although the basic two-layer hexagonal structure is maintained over the whole compositional range, we find substantial changes in the Na ion distribution as a function of layer filling in the charge reservoir layer, accompanied by significant changes in the geometry of the $CoO_2$ layer as a function of electron count, indicating the presence of strong coupling between structural and electronic degrees of freedom in this material.

**Experimental**

$Na_{0.75}CoO_2$ was made by solid state synthesis. $Na_2CO_3$ and $Co_3O_4$ in amounts corresponding to $Na_{0.77}CoO_2$ were mixed and heated to 800 degrees under flowing $O_2$ for 16 hours. Due to the volatility of Na oxides, the final Na composition is slightly lower than the starting composition. A sample of $Na_{0.34}CoO_2$ was prepared by stirring 1 g of $Na_{0.75}CoO_2$ with 1x of $NO_2PF_6$ in dry acetonitrile under argon for 2 days ("1x" indicates that the amount of $NO_2PF_6$ used is exactly the amount that would theoretically be needed to remove all the sodium from $Na_{0.7}CoO_2$). It is noted that this method produced $NaNO_3$ as a byproduct. To prepare $Na_{0.5}CoO_2$, $Na_{0.75}CoO_2$ (1 g) was stirred in 120 ml of a 10x solution of $I_2$ in acetonitrile for 4 days. $Na_{0.64}CoO_2$ was prepared by stirring $Na_{0.75}CoO_2$ (1 g) in 40 ml of a 0.5x solution of $I_2$ in acetonitrile for 4 days. The products in all cases were washed with dry acetonitrile, dried in argon, and stored an argon-filled glovebox.

For the synthesis of samples for $x > 0.75$, sodium metal (0.3 g) and tetrahydrofuran (40 mL) were placed in a 25 mm x 150 mm Pyrex screw-capped tube, and were heated by placing the lower 25 mm in an oil bath maintained at 95 deg C. After 1 h, benzophenone (2.0 g) was added. After heating another hour, powdered $Na_{0.75}CoO_2$ (1.5 g) was added, and the mixture was heated, with agitation by a magnetic stirring bar, for 48 h. The mixture was cooled to room temperature, and ethanol (20 mL) was added to destroy the benzophenone ketyl radical anion and any unreacted sodium metal. After 20 min, the mixture was centrifuged for 3 min at 2000g in a stainless steel centrifuge tube. The organic supernatant was decanted, the solid was re-suspended in dichloromethane (20 mL), the mixture was centrifuged, and the organic supernatant was again decanted. Two more dichloromethane washes were performed in the same manner, and then the final black pellet was broken up and air dried overnight to yield 1.3 g of sodium-enriched material.

This process led in two different syntheses to materials of overall stoichiometry $Na_{0.82}CoO_2$ and $Na_{0.89}CoO_2$. As described below, this difference may be due to the presence of a two-phase region in the phase diagram at these compositions, or possibly due to uncertainties in the temperature of the reaction. The sodium contents of the samples were determined by multiple independent composition measurements by the ICP-AES method, and are considered to be reliable to ±0.02/formula unit. For all samples the Na content of the constituent phases was unambiguously determined through the structural refinements.

The neutron powder diffraction intensity data for the $Na_xCoO_2$ samples were collected using the BT-1 high-resolution powder diffractometer at the NIST Center for Neutron Research, employing a Cu (311) monochromator to produce a neutron beam of wavelength 1.5403 Å. Collimators with horizontal divergences of 15´, 20´, and 7´ of arc were used before and after the monochromator, and after the sample, respectively. The intensities were measured in steps of 0.05° in the 2θ range 3°-168°. The structural parameters were refined using the program GSAS (17). The neutron scattering amplitudes used in the refinements were 0.363, 0.253, and 0.581 ($\times 10^{-12}$ cm) for Na, Co, and O, respectively.

**Results**

The structural refinements were carried out in the space group $P6_3/mmc$. The Co and O positions, and one of the possible Na positions (Na(1), in site 2$b$ (0, 0, 1/4)) were found to be unambiguously defined for all compositions. Taking the x=0.64 sample as an example, we found that the Na(2) atom, in the site 2$c$ (2/3, 1/3, 1/4), displayed a large temperature factor, ~3.0 Å$^2$, when the diffraction data were fitted with one of the models previously reported for $Na_{0.3}CoO_2$ (14). A refinement with Na(2) at the 6$h$ (2$x$, $x$, 1/4) site (labeled as the Na(2)' site), a model previously reported (15) for x=0.61, allows for displacements of the Na(2) atoms from the centers of the ideal triangular prismatic coordination polyhedra at (2/3, 1/3, 1/4). This model gave a substantially better agreement index and a more reasonable temperature factor for this Na. The possibility of a displaced position for Na(2) was therefore considered in the analysis for all compositions: if the x coordinate for the Na(2)' ions at the 6$h$ (2$x$, $x$, 1/4) site converged to x=1/3, then the Na was considered to be in the Na(2) 2$c$ (2/3, 1/3, 1/4) site. Satisfactory refinements were obtained for all compositions within these models.

The crystal structures of the three phases found in the $Na_xCoO_2$ series (table 1) are shown schematically in Figure 1. The compounds over the composition range from 0.34 < x < 0.74 (with the exception of x = 0.5) are well described by a structure in which the Na(1) site is partially occupied and the Na(2) ions are found in the positions displaced from the ideal centers of the $NaO_6$ triangular prisms, i.e. the Na(2)' sites. This structure type (15) covers a large fraction of the phase diagram, and is designated the H1 structure. Separated by a narrow two-phase region from the H1 phase, a different structure type is found for 0.76 < x < ~0.82. This phase, designated as H2, has all the Na(2) ions directly in the center of the ideal Na(2) triangular prism, as well as Na ions in the Na(1) site. The sample of overall Na stoichiometry $Na_{0.75}CoO_2$ is a mixture of the H1 and H2 phases and clearly shows the presence of the two-phase region (Figure 2).

Finally, a distinct phase (designated H3) is found at composition $NaCoO_2$ after another two phase region. In this compound, all the Na(1) sites are empty and all the Na(2) sites are filled: the Na are all in ideal trigonal prismatic sites that share only edges with the $CoO_6$ octahedra. The crystal structure of H3 is distinctly different from that of the thermodynamically stable phase $NaCoO_2$, which has all the Na in octahedral coordination with oxygen (12).

Figure 3 shows some of the structural characteristics of the $Na_xCoO_2$ phases in more detail. For Na(1), the coordination polyhedron is a regular triangular antiprism. The Na(1) position shares faces with $CoO_6$ octahedra in the planes directly above and below, a situation that leads to significant Na-Co ion repulsion, suggesting that this position is relatively higher in energy than the Na(2) and Na(2)' positions. The Na(2) position is at the center of the regular triangular antiprisms that share only edges with the $CoO_6$ octahedra. When a particular Na(2) or Na(2)' site is occupied, then its nearest neighbor Na(1) sites are excluded from occupancy, as they are too close. The Na(2)' sites are displaced from the centers of the regular antiprisms. We believe that this off-center position of the Na(2)' is found due to repulsion of the Na(2)' ion by Na ions in occupied second nearest neighbor Na(1) sites. For the H1, H2 and H3 phases (lower part of figure 3), the Co-O bondlengths are all the same within the octahedra and the O-Co-O bond angles deviate from 90º, resulting in a significant flattening of the octahedra along an axis perpendicular to the $CoO_2$ planes.

Figure 4 summarizes the compositional stability regions and general structural characteristics of the four phases found for $Na_xCoO_2$. (The O1 phase is the orthorhombic insulating phase at $Na_{0.5}CoO_2$ (6,18).) Schematic representations of the Na ion positions in the charge reservoir layer are at the top of the figure. The presence and extent of the two-phase regions (±0.02 Na per formula unit) have been determined from samples in which the overall Na content was determined by chemical analysis and two phases were observed in the neutron diffraction patterns. The fractional occupancies of the two types of sodium sites are a strong function of total Na content. The Na(1) site, where the Na ion is subject to repulsive interactions with the Co, is occupied for all compositions at a lower probability than the Na(2)' sites in the H1 phase. In the orthorhombic O1 phase (x = 0.5), however, these two sites are occupied in equal proportions, an indication of the coupling of the Na positions to the charge distribution in the underlying Co lattice (6,18). As the Na content increases from x = 0.5 towards x = 0.75, the added Na fills the Na(2)' sites only. In the H2 phase, the Na(1) site occupancy decreases dramatically with increasing Na concentration, and the Na(2) sites becomes more and more preferentially occupied. A discontinuous change in the occupancies is apparent at x ≈ 0.8: in the H3 phase all the possible Na(2) sites are occupied and the Na(1) sites are all empty.

The crystallographic *a* axis, representative of the in-plane size of the $CoO_6$ octahedra, increases in a manner consistent with decreasing the formal oxidation state of Co with increasing Na content in $Na_xCoO_2$. The shrinkage of the *c* axis lattice parameter with increasing Na content is a summation of two different effects. As the fractional occupancy of Na in the charge reservoir layer increases, the increasing positive charge results in an increasing Coulombic attraction between neighboring $CoO_2$ planes, and their separation decreases. This is shown in the upper panel of figure 5, which plots the thickness of a single $NaO_2$ layer (the $CoO_2$-$CoO_2$ plane separation) as a function of Na content. This $CoO_2$-$CoO_2$ plane separation changes continuously in the H2 phase, but there is a pronounced decrease in the H2 phase, in the composition region where all the Na(1) sites are being depopulated. This suggests that the Na(1) ion repulsion with the Co in the face shared $CoO_6$ octahedra directly above and below contributes significantly to pushing the layers apart. The overall trend to decreasing $NaO_2$ layer thickness is continued on passing from the H2 to the H3 phase.

The thickness of a single $CoO_2$ layer (Figure 5) varies in a complex manner with composition, and indicates that substantial structural changes are taking place as a response to electronic doping in $Na_xCoO_2$. In the H1 phase, the $CoO_2$ layer thickness increases systematically with increasing x, mirroring, as expected, the increasing size of the Co ion. There is a sudden increase in the $CoO_2$ layer thickness in the H2 phase, and, for $NaCoO_2$, the thickness of the $CoO_2$ layer decreases just as abruptly. Further details about the changing $CoO_2$ layers as a function of doping are shown in Figure 5. The Co-O bondlengths increase systematically with increasing electron count (decreasing formal Co oxidation state) in the H1 and O1 phases. The bond angles, however, first change in the H1 phase at low x, and then remain constant for the range of compositions $0.5 < x < 0.75$. This indicates that the $CoO_6$ octahedra do not change shape in this doping region, but rather uniformly expand. In the H2 phase there are substantial changes: both the Co-O bondlengths and the O-Co-O bond angles increase. Both of these effects lead to the observed increase of the thickness of the $CoO_2$ layer in the H2 phase within a narrow composition region. Finally, comparison of the Co-O bondlengths in the H2 and H3 phases at $x = 0.8$ and $x = 1.0$ indicates that the change in the Co-O bondlength is very small, though the degree of electronic doping of the $CoO_2$ layer has changed substantially. Surprisingly, the donation of the 0.2 electrons to the $CoO_2$ layer on passing from H2 to H3 is accommodated primarily in a change in the O-Co-O bond angle, not the Co-O bondlength.

**Discussion**

We postulate that the $CoO_2$ layer thickness is the structural characteristic that most sensitively reflects the electronic state of $Na_xCoO_2$. For $CoO_6$ octahedra of ideal shape, the $t_{2g}$ states are energetically degenerate. For octahedra compressed by shortening the distance between opposing triangular faces, as is found in $Na_xCoO_2$, the degeneracy is partially lifted, yielding two lower energy two-electron $t_{2g}$ suborbitals and one higher energy two-electron $t_{2g}$ suborbital. The greater the compression of the octahedra, the more different in energy the upper and lower $t_{2g}$ suborbitals will be. The relative thickness of the $CoO_2$ layer is a measure of that compression. Oxides based on Co3+ are well known for displaying a delicate balance between different orbital occupancies, and different spin states, based on both structural distortion and temperature (19), which affect the balance between Hund's rule coupling and crystal field effects, and in the current system itself, on hydration of $Na_{0.3}CoO_2$ to superconducting $Na_{0.3}CoO_2 \cdot 1.3H_2O$, the layer thickness decreases significantly (14) though the electron count has remained constant.

Thus there are two structural degrees of freedom to be considered when analyzing the electronic phase diagram of $Na_xCoO_2$. Our proposal for relationships between the structural and electronic properties of $Na_xCoO_2$ is summarized in Figure 6. The main panel shows the electronic phase diagram as we currently understand it (6). The upper area shows a schematic representation of the Na distribution in the hexagonal layers, the first structural degree of freedom, and the lower panel shows the ratio of the observed thickness of the $CoO_2$ layers in $Na_xCoO_2$ to that expected for layers of ideal octahedra, the second structural degree of freedom. The "ideal" octahedral layer is derived by taking the lengths of the in-plane edges of the $CoO_6$ octahedra from the measured *a* lattice parameters and employing those values to calculate an ideal layer thickness assuming 90º O-Co-O bond angles and equal Co-O bonds. We note that the $CoO_2$ layer is strongly compressed along the *c* axis: to about of 85% of the expected thickness for ideal octahedra.

The smooth variation of $CoO_2$ layer thickness with x in the H1 phase suggests that in this part of the phase diagram the effect of changing the Na content on the electronic properties of $Na_xCoO_2$, which change from superconductor host metal, through charge ordered insulator, to antiferromagnetic metal, must be a consequence of the continuously changing electron count in the $CoO_2$ plane, and not a consequence of discontinuous changes in the types of electronic states that are mixed in at the Fermi Energy. Continuous changes in the relative proportions of different types of states at $E_F$ may be present, however and mixing of the nominally empty $e_g$ states with the $t_{2g}$ and oxygen 2*p* states may be possible (20).

In contrast, H2 phase occurs in a composition region (~0.75 < x < ~0.80) of dramatic structural and electronic changes in $Na_xCoO_2$. Most significant from a structural viewpoint are the changes from occupancy of the off-center Na(2)' sites in the H1 phase to the ideally triangular prismatic Na(2) sites in the H2 phase, and the rapid depopulation of the Na(1) sites. A discontinuous change in the $CoO_2$ -$CoO_2$ layer separation is observed. This is accompanied by a substantial change in the electronic state of the $CoO_2$ layer, which rapidly increases in thickness in the H2 phase with increasing Na content. Whether these transitions are a response to the depopulation of the Na(1) sites, or are driven primarily by the change in electron count is not known, but a difference in the relative proportions of different electronic states at $E_F$ must occur. Nothing is presently known about the physical properties of materials that have been clearly identified as being in the H2 phase.

There is an abrupt decrease in the $CoO_2$ layer thickness on passing into the H2 phase, where the Na(2) sites are fully occupied in a long range ordered hexagonal lattice. $NaCoO_2$ is expected in the simplest picture ($3d^6$ $Co^{3+}$) to be a band insulator with fully occupied $t_{2g}$ orbitals. The decrease of the layer thickness on passing from the H2 to the H3 phase may reflect a change in the relative energies of filled $t_{2g}$ states compared to other states near $E_F$ at the $d^6$ electron count. The physical properties of the H3 $NaCoO_2$ phase are not currently known.

The insulating state at $Na_{0.5}CoO_2$ is found in a phase (the O1 phase) with a long-range ordered arrangement of zigzag Na chains, with Na atoms in both Na(1) and Na(2) sites, and an accompanying charge-ordered Co array. Metallic conductivity is observed in compounds with the H1 phase structure for x values surrounding x = 0.5. The displaced Na(2)' positions and the Na(1) occupancy in the average structure of the H1 phase, and the electron diffraction characterization, revealing rich variations of Na ordering schemes related to that found in $Na_{0.5}CoO_2$ (16), suggest that the Na ion array forms zigzag fragments, either disordered or in short-range ordered arrays, in the whole H1 composition region. These zigzag fragments appear to vary in number and distribution as the Na plane filling changes (16). Even in preparations or at compositions where the Na ion distribution appears to be disordered, it is likely that such fragments appear on the short-range scale, much as disordered $CuO_2$ chain fragments are found in $YBa_2Cu_3O_{7-x}$ (21). Due to the low proportion of Na(1) occupancy, such zig-zag Na fragments are not expected to occur in the H2 (or H3) phase.

In conclusion, the structural characterization of $Na_xCoO_2$ by neutron and electron diffraction has resulted in a picture of how the sodium ions are arranged in the charge reservoir layer, and how the $CoO_2$ layers respond structurally to changes in electron count and Na ion distribution. These changes are similar in character to those seen in the high Tc superconductors, where structural and electronic degrees of freedom within a square conducting plane are often linked in a subtle manner (e.g. ref. 22), suggesting that a similarly rich phenomenology remains to be fully elucidated in layered triangular conductors.

**Acknowledgements**

This work was supported by NSF grant DMR 0244254, by DOE BES grant DE-FG02-98-ER45706, and also NSF grant CHE-0314873.

Table 1. Structure parameters of Na$_x$CoO$_2$ at room temperature. Space group $P6_3/mmc$, Atomic positions: Co: 2$a$ (0 0 0); Na(1): 2$b$ (0 0 ¼) the structure types H1 & H2; Na(2): 6$h$ (2$x$, $x$, 1/4) for structure type H1 or 2$c$ (2/3, 1/3, 1/4) for structure types H2 & H3; O: 4$f$ (1/3, 2/3, $z$).

| | | | | | | | | |
|---|---|---|---|---|---|---|---|---|
| $x$ (Chem.) | nd[a] | [0.49][b] | 0.64 | 0.75 | 0.75 | nd | 0.89 | 0.89 |
| x (Refined) | 0.34 | [0.56] | 0.63 | 0.71 | 0.77 | 0.76 | 0.80 | 1 |
| Structure type | H1 | O1 | H1 | H1 (31.1%) H3 (43.1%) | H2 (68.9%) | H2 | H2 (56.9%) | |
| $a$ (Å) | 2.8129(1) | [2.81508(3)] | 2.82438(6) | 2.8314(9) | 2.84126(6) | 2.84182(4) | 2.8555(1) | 2.8829(1) |
| $c$ (Å) | 11.2152(6) | [11.1296(2)] | 11.0046(3) | 10.8756(4) | 10.8144(3) | 10.8070(3) | 10.7024(6) | 10.4927(6) |
| $V$ (Å$^3$) | 76.848(8) | [76.382(2)] | 76.024(4) | 75.867(5) | 75.606(4) | 75.583(2) | 75.576(9) | 75.524(9) |
| Co  $B$ (Å$^2$) | 0.15(7) | [0.49(4)] | 0.26(4) | 0.36(3) | 0.36(3) | 0.33(7) | 0.21(6) | 0.21(6) |
| Na(1) $B$ (Å$^2$) | 1.0 | [2.9(2)] | 1.7(2) | 1.3(1) | 1.3(1) | 1.3(2) | 1.19(7) | 1.19(7) |
|   $n$ | 0.12(1) | 0.264(9) | 0.220(9) | 0.194(19) | 0.226(10) | 0.221(12) | 0.112(12) | 0 |
| Na(2) $x$ | 0.292(4) | | 0.289(2) | 0.271(3) | | | | |
|   $B$ (Å$^2$) | 1.0 | [2.9(2)] | 1.7(2) | 1.3(1) | 1.3(1) | 1.3(2) | 1.19(7) | 1.19(7) |
|   $n$ | 0.074(4) | 0.294(8) | 0.137(7) | 0.171(9) | 0.543(13) | 0.544(16) | 0.688(14) | 1 |
| O  $z$ | 0.0855(2) | [0.08744(8)] | 0.0887(1) | 0.0904(2) | 0.0909(1) | 0.0907(2) | 0.0936(2) | 0.0936(2) |
|   $B$ (Å$^2$) | 0.80(3) | [0.66(2)] | 0.59(2) | 0.54(2) | 0.54(2) | 0.58(3) | 0.59(2) | 0.59(2) |
| $R_p$ (%) | 4.25 | [4.34] | 4.82 | 6.24 | 6.24 | 2.77 | 5.10 | 5.10 |
| $R_{wp}$ (%) | 5.26 | [5.83] | 5.74 | 5.10 | 5.10 | 3.33 | 6.46 | 6.46 |
| $\chi^2$ | 1.244 | [2.855] | 0.9538 | 0.9404 | 0.9404 | 1.015 | 1.501 | 1.501 |

Selected bond (Å) distances and angles (degrees)

| | | | | | | | | |
|---|---|---|---|---|---|---|---|---|
| Co-O  × 6 | 1.886(1) | [1.8944(5)] | 1.9004(5) | 1.911(1) | 1.9125(7) | 1.911(1) | 1.932(1) | 1.929(1) |
| Na(1)-O × 6 | 2.458(2) | [2.4321(7)] | 2.4104(8) | 2.387(2) | 2.3770(9) | 2.378(2) | 2.338(2) | |
| Na(2)-O × 4 | 2.400(2) | [2.4321(7)] | 2.345(3) | 2.300(4) | 2.3770(9) | 2.378(2) | 2.338(2) | 2.349(1) |
|   × 2 | 2.60(1) | [2.4321(7)] | 2.566(8) | 2.61(1) | 2.3770(9) | 2.378(2) | 2.338(2) | 2.349(1) |
| O-Co-O | 83.57(7) | [84.02(3)] | 84.01(4) | 84.05(8) | 84.10(4) | 83.94(8) | 83.52(8) | 83.52(7) |

[a] This sample contains 23% NaNO$_3$, it was taken into account in the refinement.
[b] Orthorhombic structure. Refinement in hexagonal symmetry for comparison purposes only. Ref. 18 reports orthorhombic refinement.

**Figure Captions**

**Figure 1.** The three hexagonal structure types found for $Na_xCoO_2$. Layers of edge-shared $CoO_6$ octahedra are seen in a triangular lattice with Na ions occupying ordered or disordered positions in the interleaving planes. Three different Na ion sites are found. In the H1 phase, $0.3 < x < 0.75$, (with the exception of $x=0.5$) Na atoms partially occupy the Na(1) $2b$ and Na(2)' $6h$ sites. In the H2 phase, $0.76 < x < \sim 0.82$, the Na(1) and Na(2) $2c$ sites are occupied. In the H3 phase, $x=1$, only the $2c$ site sites are occupied.

**Figure 2.** Observed (crosses) and calculated (solid line) intensities for $Na_{0.75}CoO_2$. Vertical lines indicate the Bragg peak positions for the phases H2 (bottom), H1(middle), and 0.4 wt% CoO impurity (top). Differences between the observed and calculated intensities are shown in the bottom of the figure. The inset shows a portion of the pattern illustrating that the (106) and (107) reflections are good indications of the different lattice parameters for the phases H2 and H3 in this sample with overall composition $Na_{0.75}CoO_2$.

**Figure 3.** Selected structural components in $Na_xCoO_2$. Upper figure: the triangular prismatic $NaO_6$ coordination polyhedra. The Na(1) ion ($2b$ site) triangular prisms share faces with the $CoO_6$ octahedra. The Na(2) ion ($2c$ site) regular triangular prisms share edges with the $CoO_6$ octahedra. The Na(2)' ions randomly occupy one of the possible the $6h$ sites distributed around the $2c$ site, displaced from the center of the triangular prisms. If a particular Na(2) or Na(2)' site is occupied, then the nearest Na(1) position is too close to be simultaneously occupied. Lower figure: detail of the $CoO_2$ layer.

**Figure 4.** The general structural characteristics and compositional stability regions of the four $Na_xCoO_2$ phases, designated as H1, H2, H3, and O1. Upper panel, hexagonal crystallographic cell parameters. Lower panel, fractional occupancies of the two types of sodium sites: Na(1) shares faces with the $CoO_6$ octahedra, and Na(2)' and Na(2) (plotted as one site) share edges with the $CoO_6$ octahedra. Above the panels, the sodium ion distributions in the four phases are shown schematically.

**Figure 5.** Upper panel, thickness of the $NaO_2$ layers (also the separation of the $CoO_2$ planes) in $Na_xCoO_2$ as a function of Na content. Middle panel: thickness of the $CoO_2$ layer and variation in O-Co-O bond angle as a function of Na content. Lower panel: the corresponding variation in Co-O bond length.

**Fig 6.** Correspondence between structure and properties in $Na_xCoO_2$. Upper figure: schematic of the Na ion distributions in the four $Na_xCoO_2$ phases. Main panel: the electronic phase diagram. Lower panel: variation of the ratio of the observed $CoO_2$ layer thickness to the thickness of an ideal layer of octahedra.

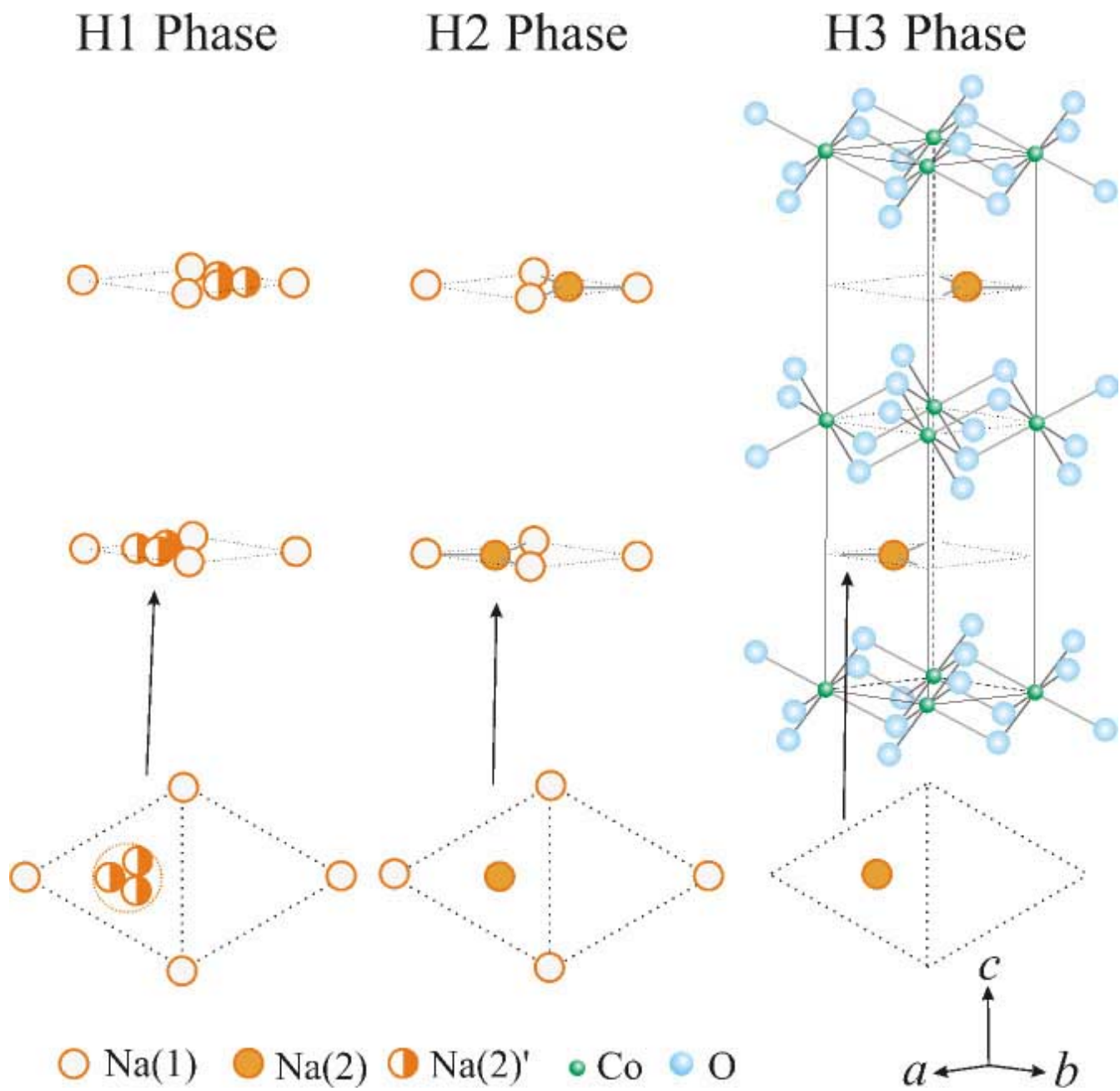

**Fig. 1.**

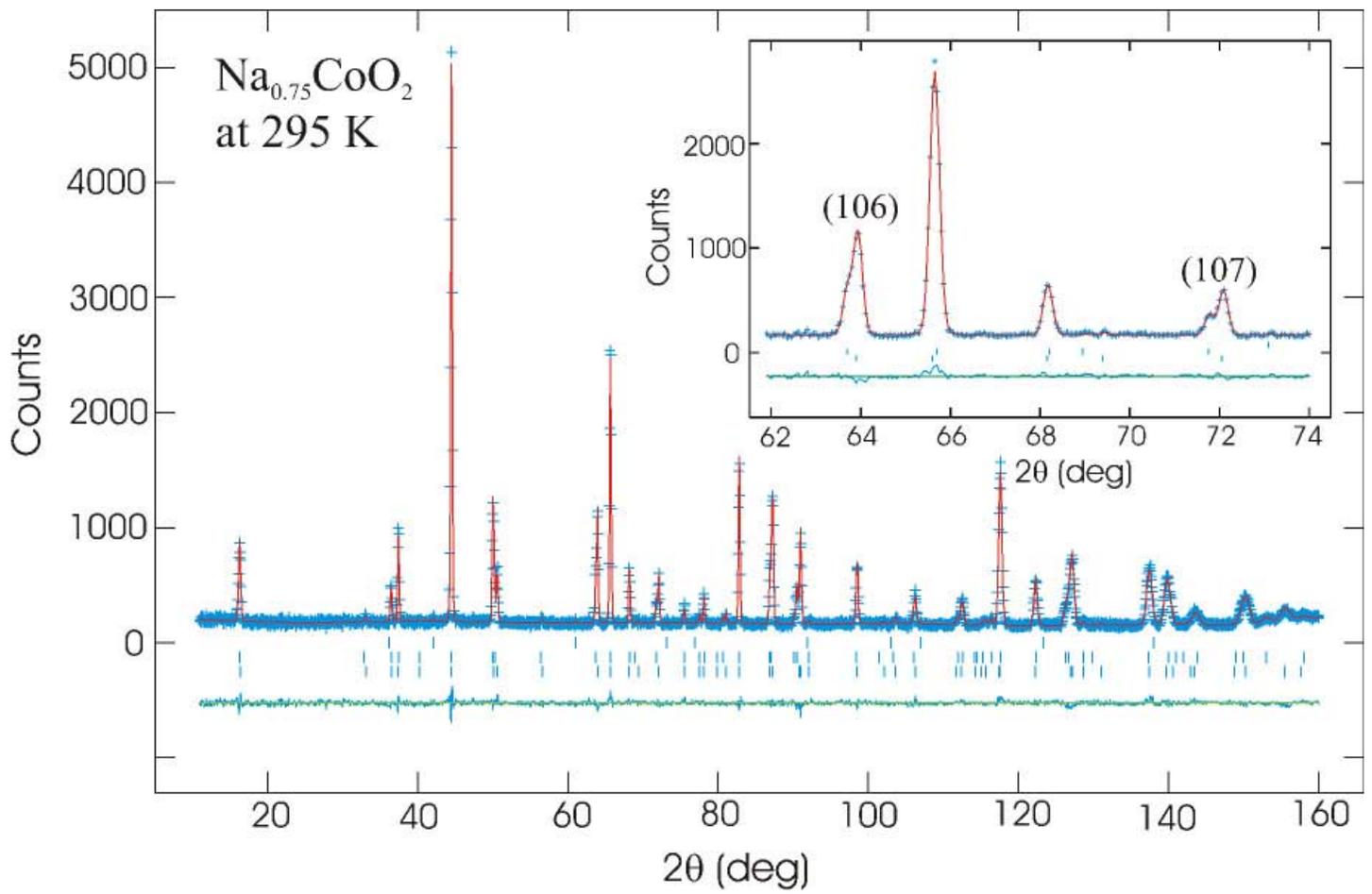

**Fig. 2.**

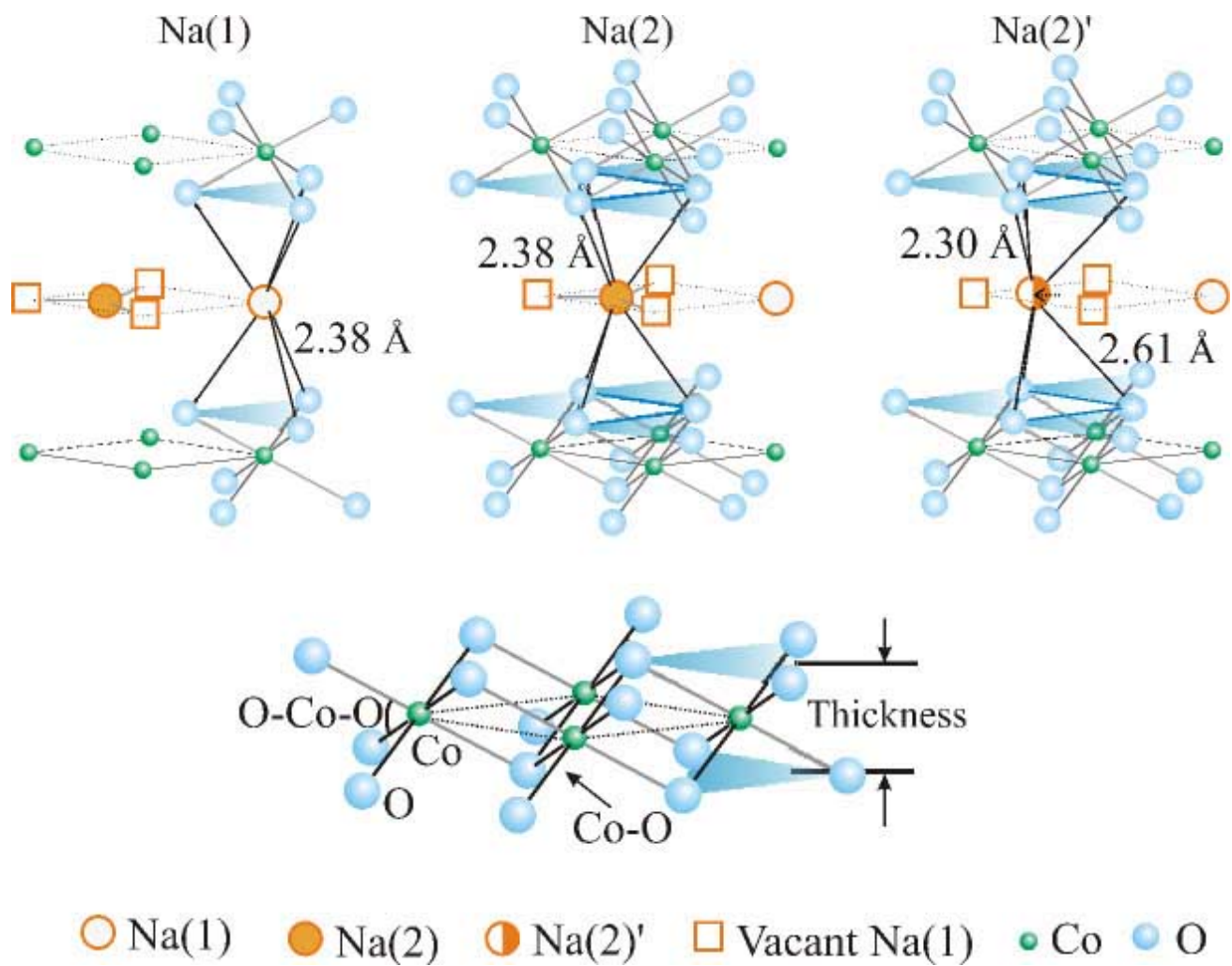

**Fig. 3.**

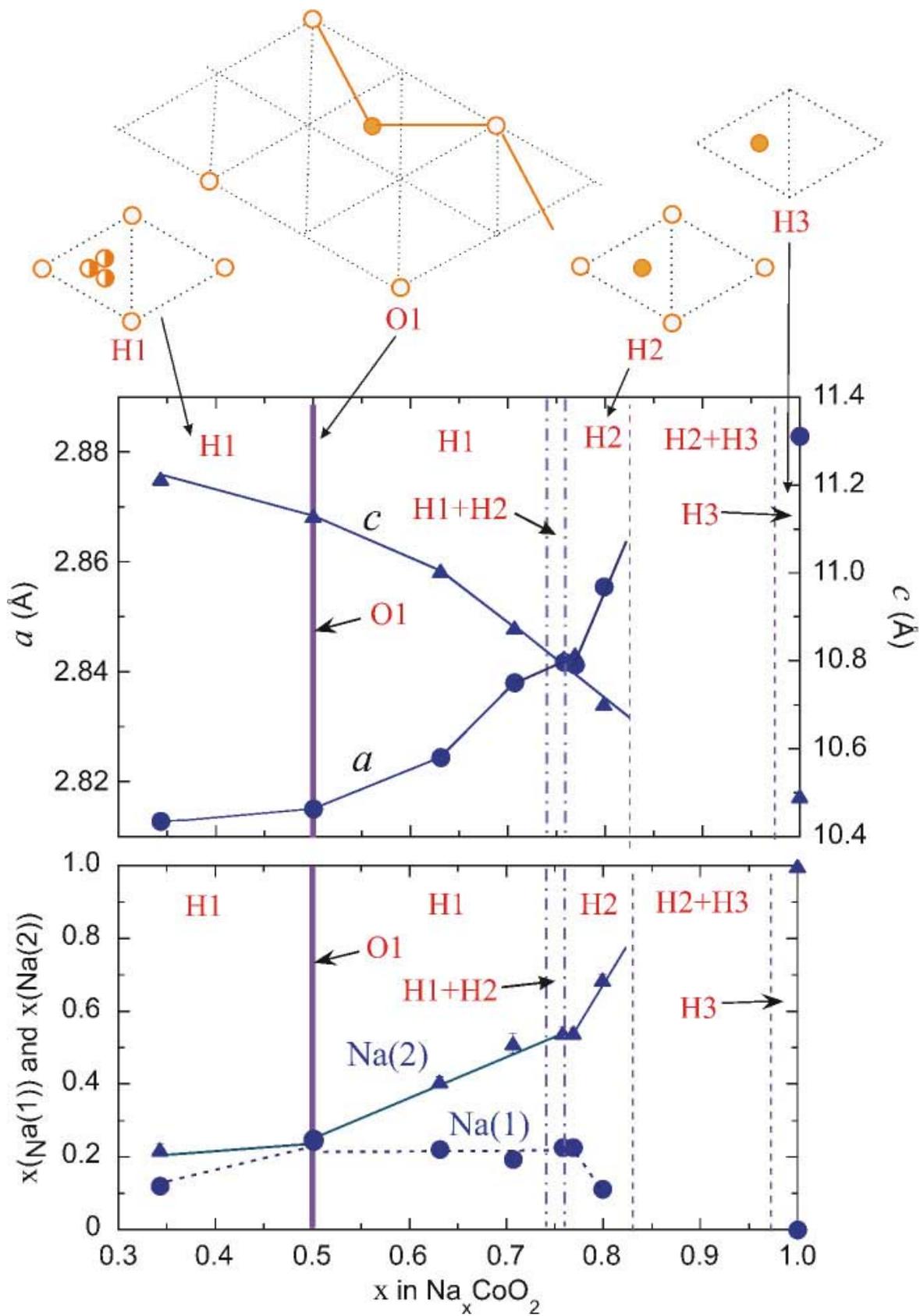

**Fig. 4**

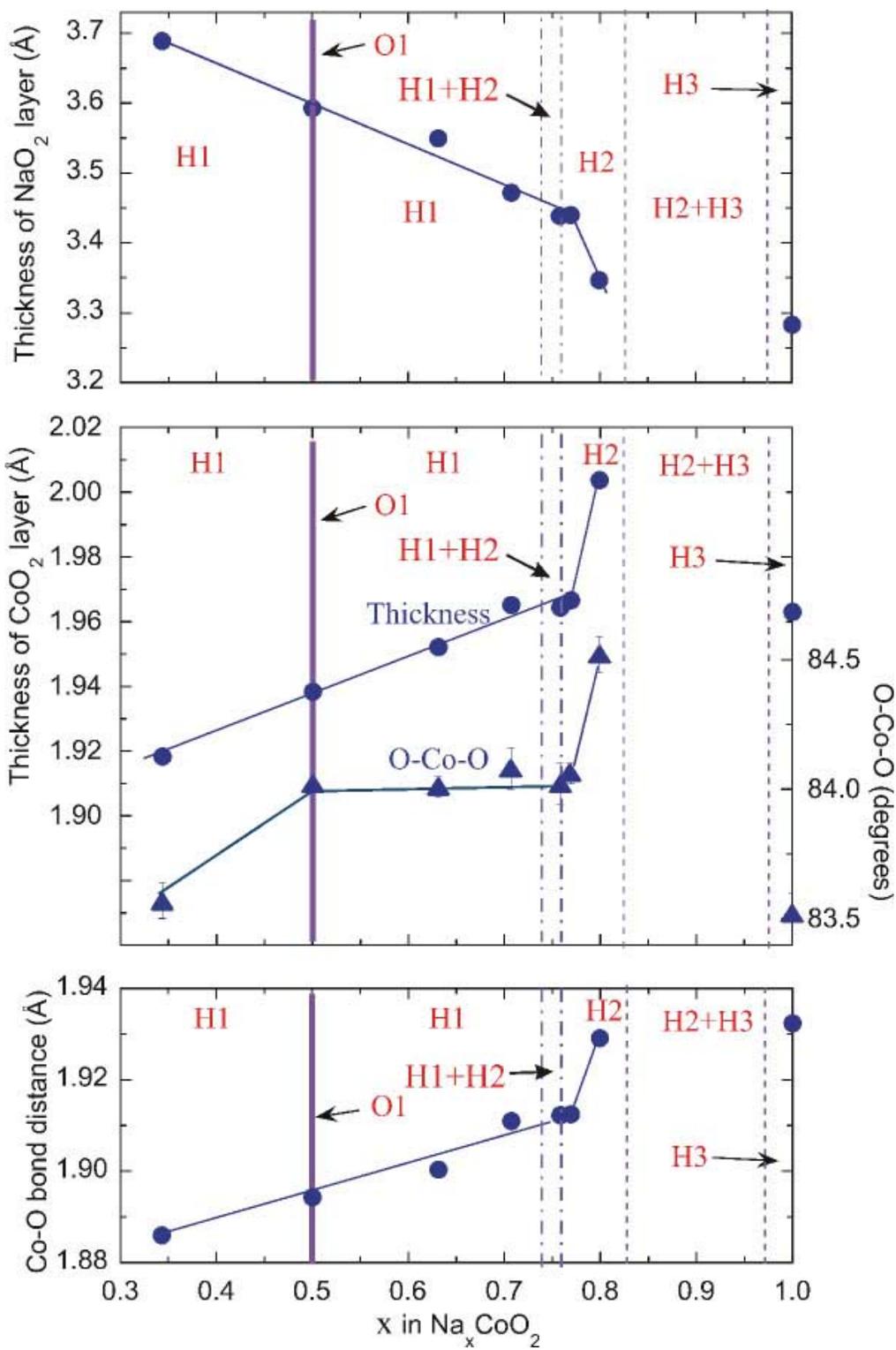

**Fig. 5**

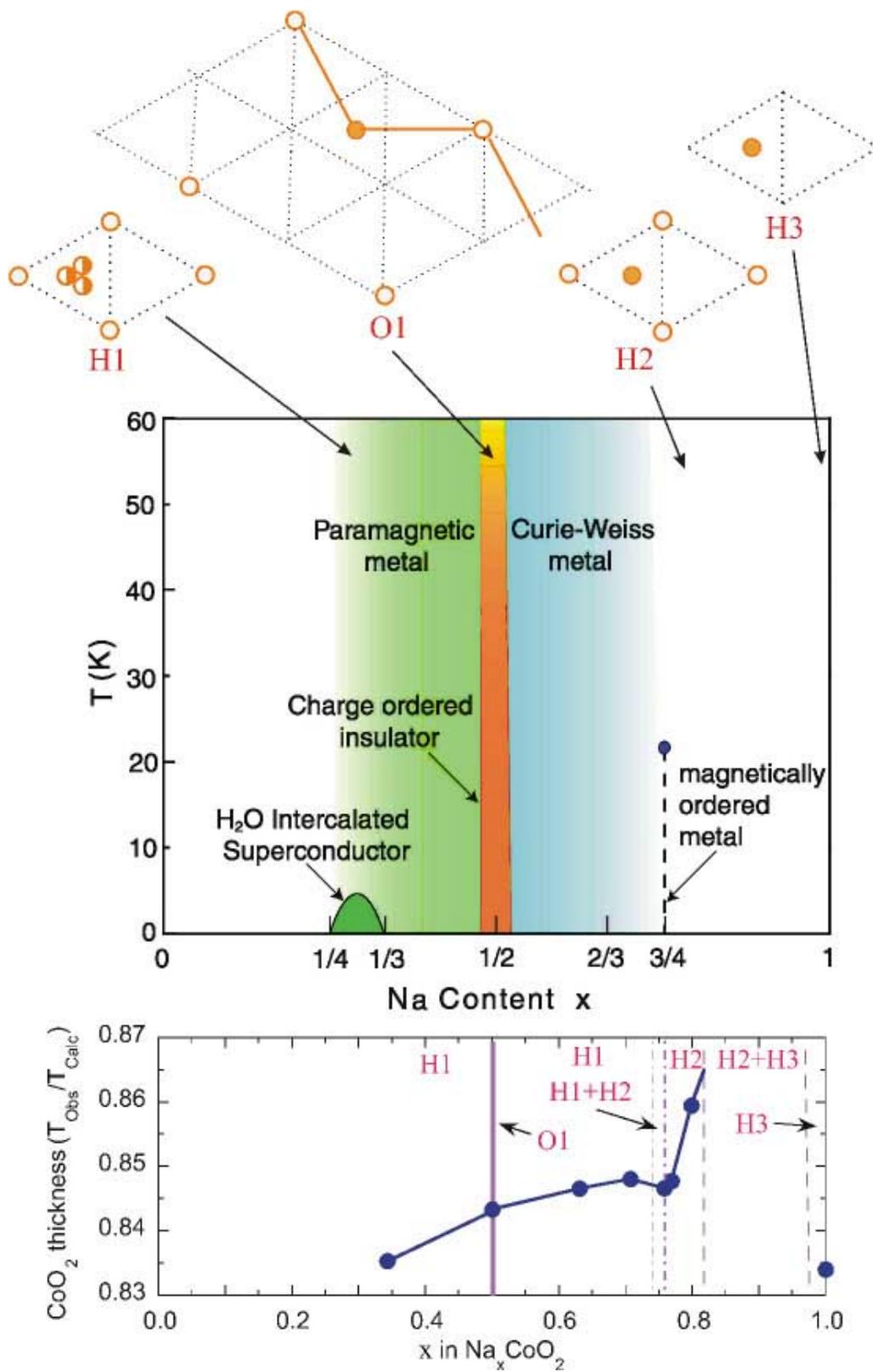

**Fig. 6**